\documentclass{article} 
\usepackage{iclr2021_conference,times}

\usepackage{amsmath,amsfonts,bm}

\def\eqref#1{equation~\ref{#1}}

\def\1{\bm{1}}

\DeclareMathAlphabet{\mathsfit}{\encodingdefault}{\sfdefault}{m}{sl}
\SetMathAlphabet{\mathsfit}{bold}{\encodingdefault}{\sfdefault}{bx}{n}

\usepackage{hyperref}
\usepackage{url}
\usepackage{graphicx}
\usepackage{caption}
\usepackage{aas_macros}

\newcommand{\bs}{\mathbf{s}}

\newcommand{\bx}{\mathbf{x}}
\newcommand{\ba}{\mathbf{a}}

\newcommand{\by}{\mathbf{y}}
\def\bi#1{\hbox{\boldmath{$#1$}}}

\title{CosmicRIM : Reconstructing Early Universe by Combining Differentiable Simulations with Recurrent Inference Machines}

\author{Chirag Modi\thanks{corresponding author}\\
Center for Computational Astrophysics,\\
Center for Computational Mathematics,\\
Flatiron Institute\\
162 5th Ave, New York, NY 10010, USA \\
\texttt{cmodi@flatironinstitute.org} 
\And
\hspace*{-35pt}Fran\c{c}ois Lanusse \\
\hspace*{-35pt}AIM, CEA, CNRS, Universit\'e Paris-Saclay \\
\hspace{-35pt}Universit\'e Paris Diderot\\
\hspace{-35pt}Sorbonne Paris Cit\'e,\\ 
\hspace{-35pt}F-91191 Gif-sur-Yvette, France\\
\hspace{-35pt}\texttt{francois.lanusse@cea.fr}
\And
Uro\v{s} Seljak \\
Department of Physics \\
University of California, Berkeley, CA, USA\\
Lawrence Berkeley National Laboratory, \\
One Cyclotron Road, Berkeley, CA, USA \\
\texttt{useljak@berkeley.edu}
\And
David N. Spergel \\
Center for Computational Astrophysics,\\
Flatiron Institute\\
162 5th Ave, New York, NY 10010, USA \\
\texttt{dspergel@flatironinstitute.org}
\AND
Laurence Perreault-Levasseur \\
Department of Physics, Université de Montréal, Montréal, Canada \\
Mila - Quebec Artificial Intelligence Institute, Montréal, Québec, Canada \\
Center for Computational Astrophysics, Flatiron Institute, New York, USA \\
\texttt{llevasseur-vscholar@flatironinstitute.org}
}

\iclrfinalcopy 
\begin{document}

\maketitle

\begin{abstract}
Reconstructing the Gaussian initial conditions at the beginning of the Universe from the survey data in a forward modeling framework is a major challenge in cosmology.
This requires solving a high dimensional inverse problem with an expensive, non-linear forward model: a cosmological N-body simulation. While intractable until recently, we propose to solve this inference problem using an automatically differentiable N-body solver, combined with a recurrent networks to learn the inference scheme and obtain the maximum-a-posteriori (MAP) estimate of the initial conditions of the Universe. 
We demonstrate using realistic cosmological observables that learnt inference is 40 times faster than traditional algorithms such as ADAM and LBFGS, which require specialized annealing schemes, and obtains solution of 
higher quality.
\end{abstract}

\section{Introduction}

Forward modeling approaches simulate the cosmological survey data, such as galaxies, in a hierarchical fashion starting from the initial conditions of the Universe. These are the most promising way of maximizing the information extraction from the next generation of cosmological surveys, because the corresponding inverse problem maps the non-Gaussian distribution of the data to a Gaussian distribution of initial conditions \citep{Seljak17}. 
This enables performing the 
marginal integral over these Gaussian initial conditions to 
obtain the data likelihood as a function of 
parameters of interest, such as the density of 
dark matter and dark energy in 
the universe and initial conditions set by inflation. 
However, to achieve this we face the challenge of solving an inverse problem in a very high-dimensional space with a complex, non-linear and expensive forward model based 
on an N-body simulation.
In this work, we demonstrate how differentiable cosmological N-body simulations, like FlowPM \citep{flowpm}, can be combined with learnt optimization methods \citep{Andrychowicz16} to learn inference schemes and tackle these challenges efficiently.
Specifically, we consider the problem of reconstructing the primordial initial conditions of the Universe from a sparse dark matter halo sample, which is a proxy for observed galaxies in cosmological surveys. 

We approach this reconstruction problem in a hierarchical Bayesian framework wherein we combine the likelihood of the observed data 
with the Gaussian prior on the initial conditions to obtain a posterior on them \citep{borg, Seljak17}. This prior depends on the initial 
power spectrum, which is the ultimate goal 
of these studies. In this work we focus on the 
reconstruction, which amounts to solving an optimization problem to obtain the maximum-a-posteriori (MAP) estimate of the 
initial conditions at a fixed fiducial 
power spectrum. 
The optimization problem is multi-million or 
even multi-billion in dimensionality, as we are interested in estimating the initial density field at all points in the Universe where we 
have observations. It is ill-conditioned 
because of noise and incompleteness in the data. 

Over the last several years, different works have built physically motivated annealing schemes relying on the dynamics of the forward models to solve this problem (\cite{Seljak17, Modi18, Modi19}). 
However this involves hundreds of iterations of the computationally expensive forward model evolving billions of particles under gravity, making the process very expensive. 
Instead, here we learn this inference scheme with a handful iterations using recurrent networks in the framework of recurrent inference machines (RIM) \citep{rim, Morningstar_2019}. 
RIM is an iterative map which includes the current reconstruction, a hidden memory state, and the gradient of the likelihood term that encodes the information about forward model, and 
proposes the next update for the current reconstruction.
RIMs were originally proposed to learn an iterative inference algorithm to solve linear inverse problems without the need to explicitly specify a prior or a particular inference procedure. 
However our motivation is different since we already know the prior on our latent parameters.
Instead, we show that with certain architectural enhancements, RIM allows us to speed up the optimization by orders of magnitude, and implicitly learn the otherwise hand-crafted annealing scheme for non-linear, ill-conditioned problems in high dimensions. 

\section{CosmicRIM}

We are interested in reconstructing the initial density field $\bx$ given a cosmological observable $\by$, which can be modeled with a non-linear differentiable forward model $\mathcal{F}$.
Let $p(\by|\bx) = N(\mathcal{F}(\bx),\bi{\sigma}^2)$ be the likelihood of the observed data, which we 
model as a Gaussian noise with variance $\bi{\sigma}^2$ that is given by the 
experiment. Let  
$p_\theta(\bx)=N(0,\bi{\theta})$ be the Gaussian prior on the initial conditions parametrized with the initial power spectrum $\bi{\theta}$, 
which we assume to be fixed at some 
fiducial value in this work.
Then our reconstruction of the initial conditions $\bx$ corresponds to the maximum-a-posteriori estimate, which 
maximized the joint $p(\bx,\by)$
\begin{equation}
    \mathrm{max_\bx} \ln p(\bx,\by) = \mathrm{max_\bx} [\ln p(\by|\bx) + \ln p_\theta(\bx) ]
\label{eq:posterior}
\end{equation}
Taking inspiration from RIM \citep{rim}, we optimize this posterior by training a recurrent neural network that follows the following update equation at every iteration $t$,
\begin{eqnarray}
\bs_{t+1} &=& \bs_t + h_\phi (\bs_t, \nabla_{x}\ln p(\bx,\by), \bx_t), \nonumber\\
    \bx_{t+1} &=& \bx_t + g_\phi (\bx_t, \nabla_{x}\ln p(\bx,\by), \bs_{t+1}).
\label{eq:rimupdate}
\end{eqnarray}
The update is informed by the gradient of the 
joint at the current position $\nabla_{x}\ln p(\bx,\by)$, as well as by the state vector $\bs_t$
that keeps track of previous information in the 
style of Recurrent Neural Networks. 
Here $g_{\phi}$ and $h_\phi$ are the neural network functions parametrized by learnt weights $\phi$. 

We construct a 10-step RIM i.e. $t=0,1\dots10$.
During training, RIM simulates all these steps of the inference and at each step the model will produce a prediction.
The total loss to train parameters $\phi$ combines all these steps to produce predictions that improve over time.
For CosmicRIM, we use a simple mean squared loss that combines every step such that 
$\mathcal{L} = \sum_{t=0}^{10} (\bx_t(\phi) - \bx_{\rm true})^2$.
For training, we use ADAM optimization  \citep{Kingma2014} with starting learning rate of $10^{-4}$.

We show architecture of CosmicRIM in Figure \ref{fig:rim}. Every convolution and LSTM kernel is of size 5 and has 16 channels. There are two more notable features of CosmicRIM - \\
\phantom{x}\hspace{1ex} 1. ADAM optimization significantly outperforms simple gradient descent for our optimization problem, with the latter failing completely.
Thus instead of using the gradients $\nabla_{x}\ln p(\bx,\by)$ directly at the current position in the update function $g_\phi$, we use the update predicted by the ADAM algorithm $\ba_t$, which itself is constructed by using the gradients from the previous RIM iterations. 
 \\
 \phantom{x}\hspace{1ex} 2. Our non-linear forward model couples different scales, and the information primarily flows from large to small scales due to gravitational evolution. Thus we use U-Net \citep{unet} like architecture with different LSTM cells \citep{lstm} in parallel to learn updates for large and small scales separately before merging them again.

\section{Results}
We consider dark matter halo field with number density $\bar{n} = 10^{-3}$ (h/Mpc)$^3$ as our observable. This will be typical of upcoming cosmological surveys like Dark Energy Spectroscopic Instrument (DESI) \citep{DESI}.
Our data is generated with $512^3$ particles evolved in  400 Mpc/h box with 40 steps of FastPM scheme \citep{Feng2016}) and force resolution B=2.

For our forward model, we evolve $64^3$ dark matter particles on a uniform periodic mesh using second order Lagrangian perturbation theory (2-LPT) and combine it with a quadratic bias model to mimic halos.
The two bias parameters, $b_1$ and $b_2$, are fit by matching the model and the data at the level of the fields in Fourier space.
This is then also used as our data likelihood in Eq. \ref{eq:posterior} with the noise variance $\bi{\sigma}^2$ given by the error power spectrum of the residual field after fitting the bias function. 
We refer the reader to \cite{Schmittfull19, Modi19}, for further details on the forward model and the corresponding likelihood function.
Figure \ref{fig:data} shows the projection of true initial conditions and the halo data as well as the recontructed fields with CosmicRIM.

\begin{minipage}[t]{0.48\textwidth}
  \centering\raisebox{\dimexpr \topskip-\height}{%
  \includegraphics[width=\textwidth]{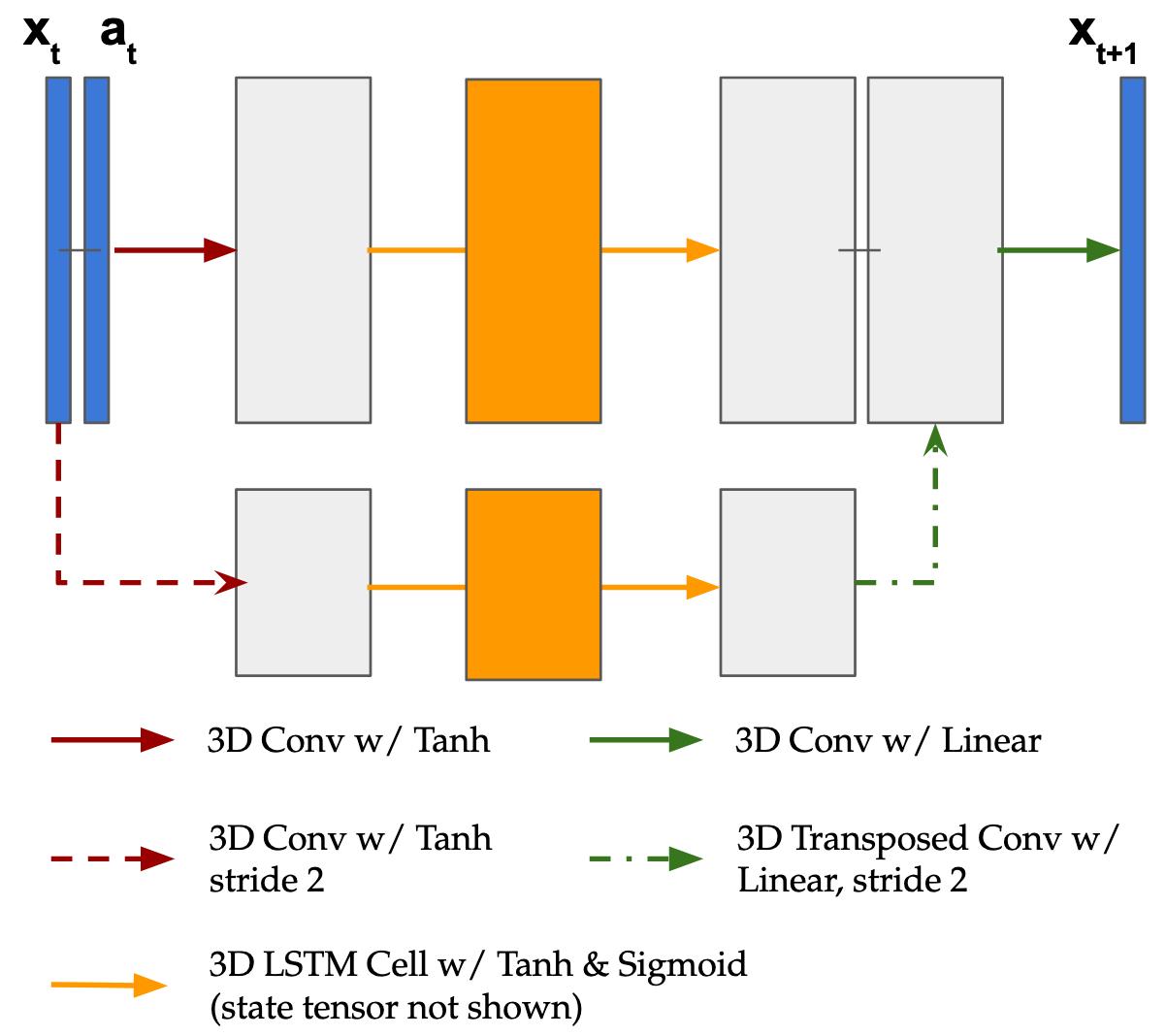}}
  \vspace{1mm}
  \captionof{figure}{Architecture of CosmicRIM: at every iteration $t$, RIM takes the current position ($\bx_t$) and proposed ADAM update ($\ba_t$) to predict an update for the next iteration ($\bx_t$)}
  \label{fig:rim}
\end{minipage}\hfill
\begin{minipage}[t]{0.48\textwidth}
  \centering\raisebox{\dimexpr \topskip-\height}{%
  \includegraphics[width=\textwidth]{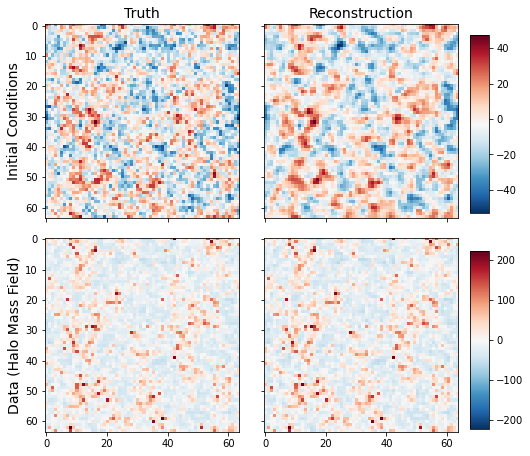}}
  \vspace{1mm}
  \captionof{figure}{Projection of the initial conditions and the halo mass data (top \& bottom row) for true fields and reconstruction with CosmicRIM (first \& second column).}
  \label{fig:data}
 \end{minipage}
 
 To statistically quantify how well we reconstruct the initial field, we compare it to the true initial field by measuring their transfer function 
 $t_f$
 , the square root of the ratio of the power spectra of two fields, 
and their cross correlation $r_{c}$
\begin{equation}
    t_f (k) = \sqrt{\frac{P_a(k)}{P_b(k)}}; \quad     r_c (k) = \frac{P_{ab}(k)}{\sqrt{P_a(k) \times P_b(k)}}
\end{equation}
with $P_{ab}$ being the cross power spectra of the two fields and $a,\ b$ correspond to reconstructed and true intitial fields in our comparison.

\begin{figure}[h]
\begin{center}
\resizebox{\columnwidth}{!}{\includegraphics{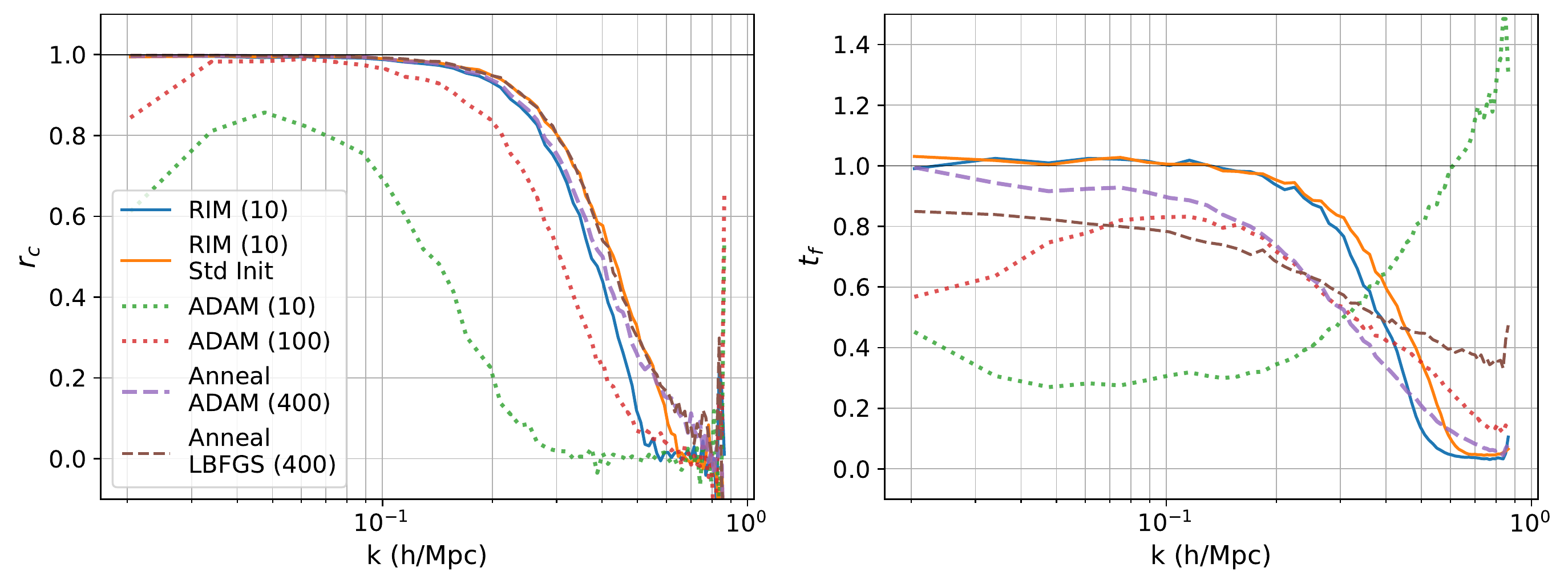}}
\end{center}
\caption{Cross correlation and transfer function of reconstructed initial field with the truth for different optimization schemes: comparing i) 10 step RIM (white noise initial position and standard reconstruction initial position) with ii) 10, 100 step ADAM and iii) annealed optimization (requires 400 steps) using ADAM and LBFGS algorithms}
\label{fig:rcc}
\end{figure}

In Figure \ref{fig:rcc}, we compare the reconstruction of CosmicRIM with other optimization schemes when starting from an un-informative white noise initial position with unit variance.
After training,	CosmicRIM takes 10 steps for reconstruction.
We compare it to the ADAM optimization for 10 and 100 steps starting from the same position and find that CosmicRIM clearly outperforms both of these. 

We also	implement an annealing scheme developed in \cite{Modi18, Feng18}, motivated by the knowledge of cosmological dynamics. The dynamics are linear on large scales and hence these should converge quickly to correct solution while information is saturated on small scales due to noise and shell-crossing. 
Hence the annealing scheme smooths the loss function on small scales to reconstruct the large scales first,	and then successively decreases	the smoothing scale.
We implement this with ADAM and	LBFGS algorithm, which is a pseudo-second order scheme, in 400 steps.
CosmicRIM is on par with these results for cross-correlation but clearly improves upon the transfer function on the large scales. However most importantly, it takes 40 times fewer iterations than these algorithms.

Performance of RIM can be significantly improved on small scales by using a more informative initial position than white noise with unit power.
We construct this initial position by doing a single step standard reconstruction \citep{Eisenstein07} with the observed data, since it takes us closest to the correct solution on the largest scales. 
Additionally, though not shown here, we verify that using ADAM or LBFGS starting from the final output of CosmicRIM does not improve our reconstruction with respect to any metric of interest, suggesting RIM already finds a good MAP for the problem. 

\begin{figure}[h]
\begin{center}
\resizebox{\columnwidth}{!}{\includegraphics{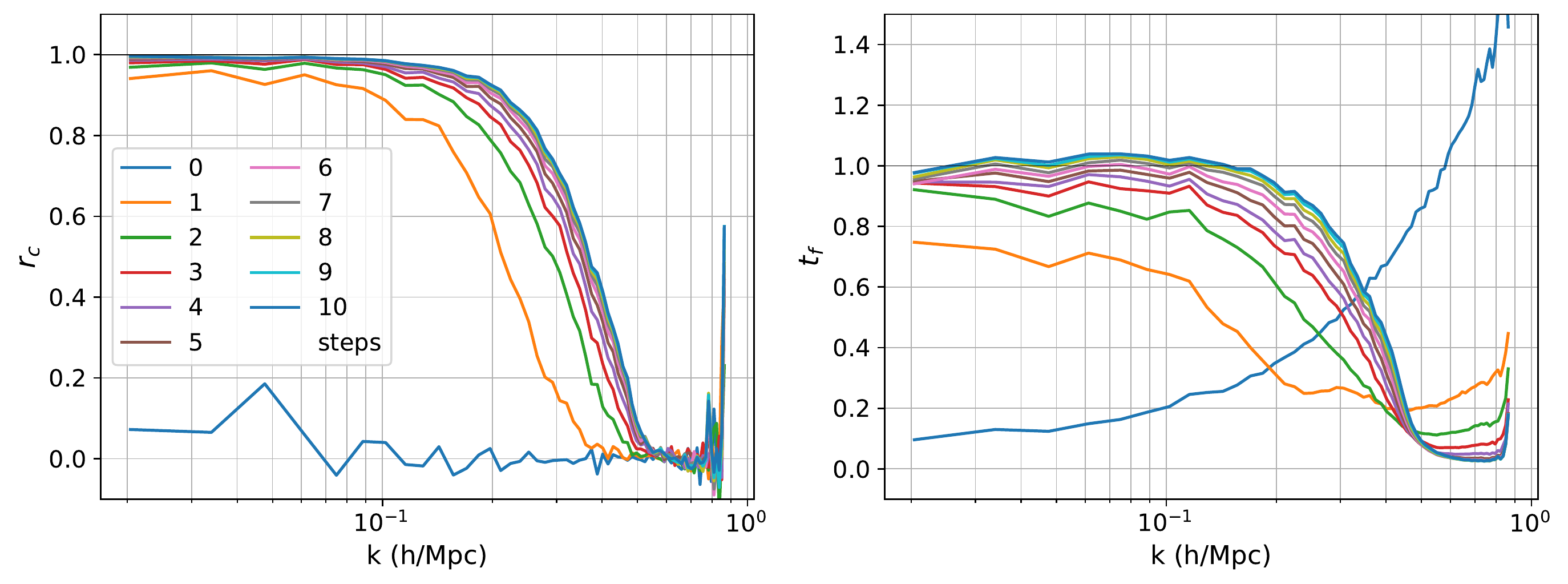}}
\end{center}
\caption{Iterations of RIM : evolution of cross correlation and transfer function over 10 steps of optimization when starting from white noise initial position}
\label{fig:niters}
\end{figure}

Finally, to understand the optimization route learnt by CosmicRIM, we also look at the metrics	for all	the 10 inference steps made by RIM	in Figure \ref{fig:niters}.
We find	that implicitly, RIM learns a scheme similar to	the one	that annealing imposes explicitly, by learning to reconstruct the more informative large scales first which are to converge faster, and then it pushes on smaller scales.

\section{Conclusion}
We combine differentiable forward models in cosmology, such as FlowPM, with learnt inference schemes such as recurrent inference machines for learning inference in high-dimensional cosmological problems with complex, non-linear forward model.
We develop CosmicRIM to reconstruct the initial conditions of the Universe from a dark matter halo mass survey data and show that it outperforms traditional optimization schemes, including domain-specialized annealing schemes, and requires 40 times fewer iterations in our tests.
Moreover, analyzing the RIM iterations, we find that it follows an optimization path similar to the one that the specialized annealing scheme is constructed to impose based on knowledge of 
cosmological dynamics. 

Fast optimization opens doors to many applications that require multiple rounds of inference on a similar class of problems. In cosmology, this can be used to reconstruct Baryon Acoustic Oscillations that involves reconstruction on a large number of mock catalogs to estimate covariance matrices \citep{Beutler16}, or for using simulations to estimate the posterior of the band powers of the reconstruced field \citep{Seljak17}.
One of the drawbacks of RIM is the high memory requirements for training. 
However in preliminary studies, we find that these can be overcome by breaking down the optimization path in steps learnt with successive networks, each trained sequentially on the output of previous networks.
This will allow to scale up the learned optimization method proposed here, as is necessary for tackling the largest of the next generation of cosmological surveys. 
\bibliography{iclr2021_conference}
\bibliographystyle{iclr2021_conference}


\end{document}